\documentclass{sig-alternate}

\usepackage{graphicx}
\usepackage{color}
\usepackage{amsmath}
\usepackage{amssymb}
\usepackage{verbatim}
\usepackage{pifont}
\usepackage{subfigure}
\usepackage{algorithm}
\usepackage[noend]{algpseudocode}
\algrenewcomment[1]{\(\triangleright\) #1}

\newcommand{\cmark}{\ding{51}}%
\newcommand{\xmark}{\ding{55}}%

\newtheorem{definition}{Definition}

\usepackage[utf8]{inputenc}  
\usepackage[T1]{fontenc}  

\usepackage{csquotes}

\newcommand{\ie}{\textit{i}.\textit{e}., }
\newcommand{\eg}{\textit{e}.\textit{g}., }
\newcommand{\cf}{\textit{c}\textit{f}. }
\newcommand{\size}{\ensuremath{\Gamma}}
\newcommand{\uacc}{\ensuremath{\theta_{acc}}}
\newcommand{\ucom}{\ensuremath{\theta _{comm}}}
\newcommand{\rem}[3]{%
\noindent{{\color{#2}{[[{\bf #1}} #3 {\bf #1}]]}}}
  
\renewcommand{\rem}[3]{}

\newcommand{\bernd}[1]{\rem{B}{red}{#1}}
\newcommand{\olivier}[1]{\rem{O}{blue}{#1}}
\newcommand{\hub}[1]{\rem{H}{green}{#1}}
\newcommand{\reb}[1]{\rem{R}{cyan}{#1}}

\begin{document}
%


\title{SPARQL Query Processing with Apache Spark}

\numberofauthors{3}
\author{
\alignauthor
Hubert Naacke\\
\affaddr{Sorbonne Universités, UPMC Univ Paris 06, UMR 7606, LIP6}\\
       \affaddr{Paris, France}\\
       \email{hubert.naacke@lip6.fr}
       \and
\alignauthor
Olivier Cur\'e\\
\affaddr{LIGM (UMR 8049), CNRS, ENPC, ESIEE, UPEM, Marne-la-Vallée, France.}\\
       \email{olivier.cure@u-pem.fr}
       \and
\alignauthor
Bernd Amann\\
\affaddr{Sorbonne Universités, UPMC Univ Paris 06, UMR 7606, LIP6}\\
       \affaddr{Paris, France}\\
       \email{bernd.amann@lip6.fr}
}

\maketitle

\olivier{
Traitement de requêtes SPARQL avec la plate-forme Apache Spark

Le nombre et la taille des graphes du Linked Open Data ne cessent de croître à un rythme soutenu et confrontent certains services sémantiques RDF à des problèmes "big data". Le traitement de requêtes SPARQL est un de ces problèmes et doit être considéré de manière attentive pour proposer des systèmes avec des garanties de montée en charge, de  haute disponibilité et de tolérance aux pannes. Ces propriétés sont complexes et souvent réalisées avec beaucoup d'effort dans des plateformes géneriques et ouvertes à des nombreuses applications. Dans cet article, nous considérons le traitement de requêtes SPARQL avec Apache Spark, une plateforme récente et représentative pour le traitement massif de données. Nous proposons et comparons cinq approches d'évaluation de motifs SPARQL différentes fondées sur deux algorithmes de jointures distribuées classiques (jointure par partitionnement et jointure par réplication) et sur les différentes couches de stockage Spark. Une expérimentation détaillée sur des jeux de données réelles et synthétiques met en évidence que les approches hybrides mélangeant les deux types d'algorithmes de jointures sont les plus performantes.}

\begin{abstract}
The number and the size of linked open data graphs keep growing at a fast pace and confronts semantic RDF services  with problems characterized as Big data. Distributed query processing is one of them and needs to be efficiently addressed with execution guaranteeing scalability, high availability and fault tolerance. RDF data management systems requiring these properties are rarely built from scratch but are rather designed on top of an existing engine. In this work, we consider the processing of SPARQL queries with the current state of the art cluster computing engine, namely Apache Spark. 
We propose and compare five different query processing approaches based on different join execution models and Spark components.  A detailed experimentation on real-world and synthetic data sets promotes two new approaches tailored for the RDF data model which outperform (by a factor of up to 2.4 on query execution time compared to a state of the art distributed SPARQL processing engine) the other ones on all major query shapes, \ie star, snowflake, chain and their composition.
\end{abstract}





\terms{Distributed Query Processing, Graph Partitioning, Distributed Join Algorithms, Map-Reduce}


\section{Introduction}

\hub{Cibler assez rapidement le cas d'usage d'un workflow composé de 2 types de tâches sur un grand graphe RDF: analyse globale de la structure (exple calculer un page rank ou le degre des noeuds) et requêtes ciblées sur un motif complexe (etoile, chaine, flocon)

Besoin de composer des tâches qqsoit leur types (analyse et/ou requête) SANS exporter/importer les données: toutes les tâches doivent partager les données dans un espace mémoire commun.}

The Semantic Web is growing very rapidly and generates large volumes of RDF data~\cite{rdf11}. 
Tasks like semantic knowledge discovery and data integration generally rely on the access to several billions of RDF triples contained in the Linked Open Data (LOD) cloud or other open data and knowledge sources like Schema.org.
RDF data management  is a central component of these tasks. With data sets ranging from hundreds of millions to billions of RDF triples, RDF data management is expected to meet properties such as scalability, high availability, automatic work distribution and fault tolerance. 
Supporting efficient SPARQL query processing in such a context becomes an important challenge and
we are convinced that without robust and efficient RDF stores, the success of the Semantic Web vision is at stake.


The features expected from modern RDF stores are reminiscent of the Big data trend where solutions implementing  specialized data stores from scratch are rare due to the enormous amount of development effort they require.  Instead, many RDF data processing systems prefer to rely on an existing cluster computing engine based on the MapReduce approach \cite{DBLP:conf/osdi/DeanG04} and massively parallelized  distributed data processing in general.  Nevertheless, these cluster engines cannot be considered as full-fledged data management systems \cite{DBLP:journals/cacm/StonebrakerADMPPR10} and  integrating an efficient query processor on top of them is already a challenging task. In particular, data storage and communication costs generated by the evaluation of joins over distributed data need to be addressed cautiously.  Furthermore, new data sets of ever increasing size need fast loading and storage strategies to shorten data ingestion time before querying. Thus, we often have to trade data-to-query storage locality for better timeliness.  

In a cluster-based setting, both the data and the query processing are highly distributed. As a consequence, complex SPARQL queries over large RDF graphs generally have to combine a lot of distributed pieces of data through join operations.  Choosing the optimal data distribution and join strategy that minimizes data transfer is difficult because it  depends on \emph{a priori} knowledge about the data and/or query workload which is often difficult to obtain under big data assumptions. To compensate the absence of this knowledge one needs a cautious just-in-time investigation among various ways to evaluate joins.
%
Several systems addressed these problems using the MapReduce cluster computing approach~\cite{DBLP:conf/oopsla/RohloffS10,DBLP:journals/pvldb/HuangAR11,DBLP:journals/pvldb/LeeL13,DBLP:conf/icde/GoasdoueKMQZ15}. Nonetheless, these systems did not reach expected performances due to the inadequacy of MapReduce for tackling interactive or iterative tasks that we are aiming for in a full-fledged distributed RDF store. The main reason is its lack of an abstraction for efficiently handling the main memory distributed over the cluster's machines.
Nevertheless, CliqueSquare\cite{DBLP:conf/icde/GoasdoueKMQZ15}  is a MapReduce based system that is worth mentioning due its effort to maximize local joins. But the proposed solution replicates the whole data set three times which is not applicable to a main-memory approach.

Recent systems, such as Apache Spark~\cite{DBLP:conf/hotcloud/ZahariaCFSS10} or Apache Flink\footnote{https://flink.apache.org/index.html} have been designed to address these drawbacks.
To the best of our knowledge, S2RDF~\cite{s2rdf} is the only, with the prototype presented in this paper, system that processes SPARQL queries with Apache Spark. Nevertheless, the approaches are different: S2RDF proposes a novel relational schema (denoted extVP) and relies on a translation of SPARQL queries into SQL for being executed using Spark SQL. In this work, we show that
Spark SQL does not (yet) fully exploit the variety of distributed join algorithms and  plans that could be executed using the Spark platform and propose some first directions for more efficient implementations.

\reb{
Motivation and related work (Rev3) : we wanted to compare different distributed query rewriting/processing strategies (logical/physical rewriting, …) within a uniform setting. The choice of the generic Spark platform (Rev4) allowed us to compare different strategies within a uniform software environment eliminating low-level performance side-effects related to using different technologies. This is a first step for defining more general optimization rules and frameworks which could be integrated into other systems. Nevertheless, we agree that a deeper comparison with other distributed RDF graph query processing systems and systems like Impala is necessary as future work.
}

The contributions presented in this article are the following: (1) we propose a formalization for evaluating the cost of SPARQL query processing in a distributed setting (Section~\ref{sec:sparql}), (2) we design and compare five Spark-based SPARQL query processing solutions using our theoretical framework (Sections~\ref{sec:query:spark} ), (3) we finally validate our framework and evaluate these solutions by an experimental evaluation over real-world and synthetic data sets (Section~\ref{eval}). To the best of our knowledge, this is the first theoretical and experimental evaluation of Apache Spark for processing SPARQL queries at this level of detail.

\section{Distributed SPARQL Processing}
\label{sec:sparql}

In this section, we first briefly introduce RDF and the SPARQL query
model.
We present the usual horizontal data partitioning techniques that suit for distributing RDF large data sets.
Then, we introduce a simple physical algebra for the distributed evaluation of SPARQL query expressions, \ie \emph{triple selections} and n-ary \emph{triple joins}.
Based on this algebra, we present a cost model estimating data transfers, and two commonly used distributed join processing~\cite{DBLP:conf/sigmod/BlanasPERST10} methods, \ie partitioned join and broadcast join.
\bernd{j'ai ajouté un peu de précisions dans l'explication des coûts: la mesure concernant la taille n'était pas claire. Pour simplifier (?) je propose que la taille s'exprime en nombre de triplets et les coûts d'accès et de transfert prennent en compte les différence de compression de RDD et DF; à vérifier}

\subsection{RDF and SPARQL}
\label{sec:rdf}

RDF is a schema-free data model that supports the description of data on the Web. 
Assuming disjoint infinite sets I (RDF IRI references), B (blank nodes) and L (literals), a triple (s,p,o) $\in$ (I $\cup$ B) x I x (I $\cup$ B $\cup$ L) is called an RDF triple with s, p and o 
respectively the subject, predicate and object. 
Assume that V is an infinite set of variables and that it is disjoint with I, B and L. 
We can recursively define a SPARQL\footnote{http://www.w3.org/TR/rdf-sparql-query/} triple pattern as follows: 
(i) a triple $tp \in$ (I $\cup$ V) x (I $\cup$ V) x (I $\cup$ V $\cup$ L) is a (simple) triple pattern, (ii) if $tp_1$ and $tp_2$ are triple patterns, then $tp_1 . tp_2$ represents a group of triple patterns 
that must all match, ($tp_1$ \texttt{OPTIONAL} $tp_2$) where $tp_2$ is a set of patterns that may extend the solution induced by $tp_1$, and ($tp_1$ \texttt{UNION} $tp_2$), denoting pattern alternatives, are 
triple patterns and (iii) if $tp$ is a triple pattern and C is a built-in condition then the expression ($tp$  \texttt{FILTER} C) is a triple pattern 
that enables to restrict the solutions of a triple pattern match according to the expression C. 
The SPARQL syntax follows the select-from-where approach of SQL queries. The \texttt{SELECT} clause specifies the variables appearing in the result set of the query, the \texttt{FROM} clause specifies the data sets to be used for matching and the \texttt{WHERE} clause defines the triple pattern and optional sub-queries. 

\subsection{Pattern queries}
In this article, we are focusing on the evaluation of groups
(sets) of simple triple pattern expressions without filters,
alternatives and union. 
Efficiently evaluating such patterns is essential for all SPARQL query engines and one of the most important challenges in SPARQL query optimization.
\begin{definition}
\label{def:query}
Let $P=\{t_1, \cdots t_n\}$ be a group pattern (set of triple patterns) with variables $V$. 
We will define \emph{query plans} $Q$ for $P$ as a composition of join expressions:
\begin{itemize}
\item Any triple pattern $t_i$ with a set of variables $V_i$ is a (triple selection) query (with variables $V_i$).
\item Let $q_1, \cdots, q_n$ be queries with variables $V_1, \cdots, V_n$ and $V=V_1 \cap \cdots \cap V_n$ be the set of all variables shared by all queries $q_i, 1\leq i\leq n$, then $join_V(q_1, \cdots, q_n)$ is a (n-ary join) query with variables $V'=(V_1 \cup \cdots \cup V_n)$.
\end{itemize}
A query $Q$ is a (logical) query plan for some pattern $P$ iff $Q$
contains exactly the set of triple patterns of $P$.
\end{definition}


\begin{figure*}[htbp]
\centering
\includegraphics[height=3.6cm]{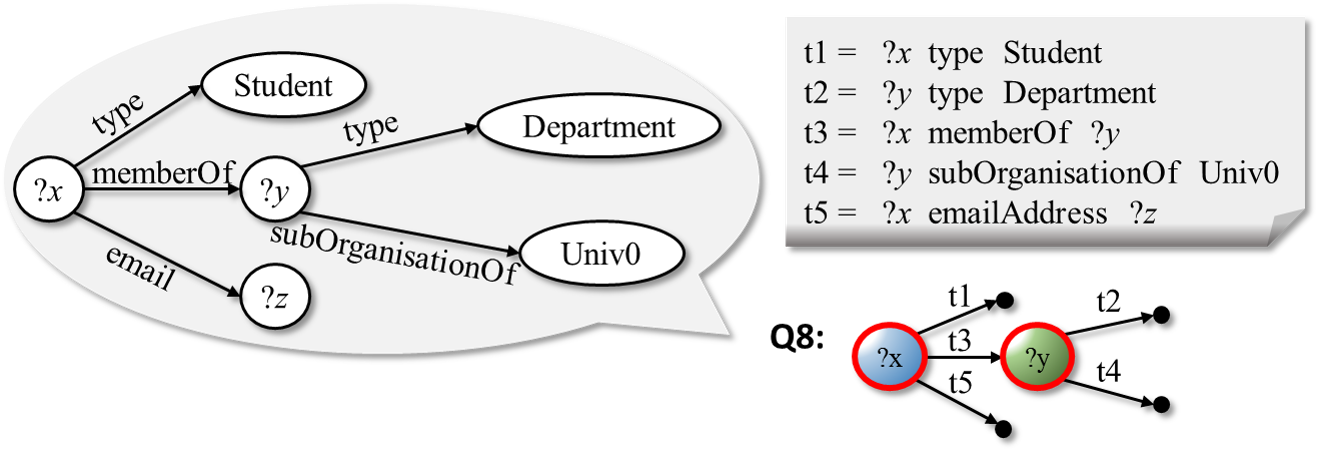}
\caption{Q8 query of the LUBM benchmark}
\label{fig:q8}
\end{figure*}
Figure~\ref{fig:q8} shows a group pattern $P$ aiming to retrieve the email address $?z$ of all students $?x$
who are members of department $?y$ in university $Univ0$. Each triple
pattern $t1, ...,t5$ implicitly defines a \emph{triple selection}
which computes all triples respecting this pattern. For example $t4$
filters all triples with property $subOrganisationOf$ and object
$Univ0$ and \emph{binds} variable $?y$ to the subjects of these
triples. Variables $?x$ and $?y$ are called \emph{join variables},
since they define n-ary \emph{triple pattern join}s:
$join_x(t_1,t_3,t_5)$ and $join_y(t_3,t_2,t_4)$ where the first join
expression joins all triples on their subject, whereas the second
expression, joins the object of $t_3$ with the subjects of $t_2$ and
$t_4$.  The first query plan for $P$ is then $Q_{81} = join_x(t1,
join_y(t_3,t_2,t_4), t_5)$ and the final result is the \emph{set of
  bindings for $?x$, $?y$ and $?z$} obtained after the evaluation of
all selection and join operations. Observe also that we do not impose
any ordering on the selection and join operations at the logical
level
and that it is possible to build several equivalent plans for $P$. For
example, the two $t_i$ candidates that can join with $t_4$ are $t_2$
and
$t_3$ 
and it is possible to decompose the n-ary joins into several binary
joins to obtain query $Q_{84} =
join_x(join_x(join_y(join_y(t_4,t_2), t_3), t_1), t_5)$.

Our definition also allows to characterize the notions of star, chain, snowflake and complex patterns. A pattern $P=\{t_1, \cdots t_n\}$ is a (1) star, (2) chain or (3) snowflake pattern, if there exists an equivalent query plan $Q$ where 
\begin{description}
\item[star:] $Q=join_x(t_1,\cdots,t_n)$ where $x$ is a single variable at the subject or the object position in $t_i$. $Q$ is called oriented if $x$ is at the same position (subject or object) in all $t_i$.
\item[chain:] $Q=join_{x_1}(\pi_1, join_{x_2}(\pi_2, \cdots join_{x_{n-1}} (\pi_{n-1},\pi_n)\cdots))$  where all $x_i$ are different single variables at the subject or object positions in $t_i$ and $\pi$ is a permutation over all triples in $P$.
\item[snowflake:]  $Q=join_{x_1}(P_1, join_{x_2}(P_2, \cdots join_{x_k} (P_k)\cdots))$ where all $x_i$ are different single variables at the subject or object positions in $t_i$  and $P_i$ are non-empty partitions of $Q$ (by definition $P_k$ contains at least 2 triple patterns). 
\end{description}
All other queries are called complex queries.

\bernd{j'ai ajouté un peu de précisions dans l'explication des coûts: la mesure concernant la taille n'était pas claire. Pour simplifier (?) je propose que la taille s'exprime en nombre de triplets et les coûts d'accès et de transfert prennent en compte les différence de compression de RDD et DF; à vérifier}


In the following we will introduce a simple distributed algebra and
cost model for evaluating query expression as defined above. Given a data set $D$, a query expression $Q$ and a cluster of nodes $C$, the
global query evaluation process is as follows:
\begin{itemize}
\item The data set is partitioned and distributed once over the cluster following a predefined query-independent hash-based partitioning strategy described below. 
\item Each node can evaluate any triple selection locally over its own triple set.
\item The join operations are recursively executed following a distributed physical join plan using different physical join implementations. The definition and comparison of different representative join plans is one major contribution of this article.
\end{itemize}

In the following we will describe each step in more detail.
\subsection{Data partitioning}
\label{sec:datapart}

\reb{ Data partitioning (Rev3): since our approach is main-memory oriented and, although data are retrieved from a persistent layer (such as HDFS which uses its own partitioning approach), we partition the data set based on hashing on triple subjects (using Spark’s native partitioner). The partitioning strategy used at the persistent storage layer is orthogonal to query processing because we consider that the whole data set fits in main-memory}

Due to its high efficiency, hash-based partitioning is the foundation of MapReduce-based parallel data processing infrastructures. Consider a cluster $C=(node_1, \cdots node_m)$  of $m$ nodes and some query $q$ with variables $V$ over an input data set $D$. A \emph{hash partitioning strategy} for $q$,  denoted $q^{V'}$, consists in partitioning the result of $q$ with respect to the bindings of a variable subset $V'\subseteq V$. 

For example, $(?x~prop~?y)^{?x}$ denotes that all triples with property $prop$ are partitioned by their subject, $(?x~?p~?y)^{?p}$ denotes a vertical partitioning by property type and $(?x~?p~?y)^{?x}$ denotes a  horizontal partitioning by subject. It is also possible to partition by subject and object, $(?x~?p~?y)^{?x~?y}$. 

By definition $(?x~?p~?y)^{?x~?p~?y}$ is equivalent to $(?x~?p~?y)^{\emptyset}$ and denotes a random partitioning. In the following, we suppose that all triples of the input data are partitioned by their subject.

\subsection{Triple Selection}
\label{sec:select}
Given a triple pattern $t$, the triple selection algorithm consists in scanning the whole input data set $D$ (no indexing assumption) as detailed in Algorithm~\ref{alg:sel} where $\pi$ and $\sigma$ denote the relational projection and selection operator over a table with three attributes $s$,  $p$  and $o$.
\begin{algorithm}
\caption{Triple selection} 
\label{alg:sel}
\begin{algorithmic}[1]
  \State \textbf{Input}: $t:(x, y, z)$, $D$
\State \textbf{Output}: result fragment $Result_j$ on each node $node_j$
  \Comment{Build query expression}
  \State $proj \leftarrow \emptyset$
  \State $cond \leftarrow true$
  \If {$x$ is a variable}
    \State $proj \leftarrow proj \cup \{x\}$
    \Else { $cond \leftarrow cond \land (s = x) $}
  \EndIf
  \If {$y$ is a variable} 
    \State $proj \leftarrow proj \cup \{y\}$
    \Else { $cond \leftarrow cond \land (p = y) $}
  \EndIf
  \If {$z$ is a variable}
    \State $proj \leftarrow proj \cup \{z\}$
    \Else { $cond \leftarrow cond \land (o = z) $}
  \EndIf
  \Comment{Compute the variable bindings matching $t$ (locally)}
  \ForAll {$node_j $ } 
    \State $Result_j \leftarrow \pi_{proj}(\sigma_{cond}(d_j))$
  \EndFor
\end{algorithmic}
\end{algorithm}
All triple selections are evaluated locally on each cluster node and generate no data transfer.
The cost to evaluate $t$ only depends on the size $\size(D)$ (in
kilo-bytes) of a given data set $D$ and the unit data access cost
$\uacc$.  Let $d_{j}$ denote the chunk of $D$ on $node_j$ in cluster
$C$. Then the cost of evaluating a single triple pattern $t$ is


$$Cost(t) = \uacc \times \size(D) = \uacc \times \sum_{i=1}^{m}  \size(d_i)$$

The triple selection preserves the data partitioning, \ie the result
of a triple selection has the same partitioning property as the input
data set. For instance, considering the $Q_8$ query over a triple set $D$ partitioned by subject, we would have $D^s$, $t_1^x$, $t_3^x$, $t_5^x$,
$t_2^y$, and $t_4^y$.

\bernd{check the following:
 For the sake of simplicity, we assume also that all triple patterns are selective and the result size of a triple selection is rather small with respect to the size of the input data set.\footnote{This assumption is true for most patterns except certain patterns using property \texttt{rdf:type}, like for example \texttt{(?x rdf:type rdf:Resource)} which returns the whole data set.}
}

\paragraph{Merging multiple triple selections} 
\label{sec:multisel}
The goal of this triple selection approach is to minimize the access cost by scanning in a first pass the minimal data relevant to several  sub-expressions of a join query. 
%
%
Let $(t_1, \cdots, t_n)$ be the set of triple patterns of a query $q$. Then, the total access cost for all triples in  $q$ is
$$\sum\limits_{i=1}^{n} Cost(t_i) = n \cdot \uacc \cdot \size(D)$$

%
%
%

%

Since all $t_i$ are expressed over the same data set $D$, there are
opportunities to save on access cost for evaluating
$q$. 
The basic idea is to replace $n$ scans over the whole data set $D$ by
a single scan over the whole data set and $k$ scans over a much
smaller sub-set.
For this, we first rewrite the selections in $q$ into a single
selection $S=\sigma_{c_1 \lor \cdots \lor c_n}$ where $c_i$ is the select condition of $t_i$
which returns all triples $\bigcup_{i=1}^{n}t_i$ necessary for
evaluating $q$. 
The query access cost becomes $$\sum\limits_{i=1}^{n} Cost(t_i) =
\uacc \cdot (\size(D) + n \cdot \size(S))$$
Merging multiple selections reduces the data access
cost if the following condition holds iff 
$\uacc \cdot (\size(D) + n \cdot \size(S)) < n \cdot \uacc \cdot \size(D)$ or $$\size(D)/\size(S) <
\frac{n}{n-1}$$

%





Our experiments will show the benefit of merging multiple triple selections on star pattern queries, and on other queries with more complex query shapes (like snowflake or chain queries).



\subsection{Partitioned Join: $Pjoin$} 
\label{sec:pjoin}
Let $q=join_V(q_1^{p_1}, \cdots, q_n^{p_n})$ be an n-ary join query as
defined in Section~\ref{sec:datapart}.   The
principle of the $Pjoin$ operator, denoted $Pjoin_V(q_1^{p_1},
\cdots, q_n^{p_n})$, is to partition (when necessary) the input data over the bindings of all variables in 
$V$ and to compute the result independently (and possibly in
parallel) for each partition.
Let $d_i$ be the result of $q_i$, and $d_{ij}$ be the chunk of $d_i$
on $node_j$ ($1\leq j\leq m$).  The partitioned join algorithm evaluates the query
result 
in four steps as detailed in Algorithm~\ref{alg:partj}.  After reading
the input set $d_i$ of each sub-query (lines 3-5), partition (if necessary) each $d_i$
based on the join key $V$ such that the triples having the same value
for $V$ into $m$ partitions (lines 6-7). The third step transfers
(\ie shuffles) the data to the $m$ target nodes (lines 8-9) such that
each node is responsible to compute the join for some values of $V$
(lines 10-11). The result of $q$ is partitioned on $V$, which is
denoted $q^V$.






\begin{algorithm}
\caption{Partitioned Join} 
\label{alg:partj}
\begin{algorithmic}[1]
\State \textbf{Input}: $\{q_1^{p_1}, \cdots, q_n^{p_n}\}$, join variables $V$
\State \textbf{Output}: result fragment $Result_j$ on each node $node_j$
\Statex \Comment{Evaluate, partition and shuffle sub-query results}
\ForAll{$q_i$} 
  \ForAll{$node_j$} 
  \State $d_{ij} \leftarrow$ evaluate $q_{i}$ on node $node_j$
  \If{$p_i \neq V$}
    \State partition $d_{ij}$ on $V$ into $\{d_{ij1}, \cdots, d_{ijm} \}$
       \ForAll{$node_k \neq node_j$ }  
          \State transfer partition $d_{ijk}$ from $node_j$ to $node_k$
       \EndFor
  \EndIf
  \EndFor
\EndFor
\Statex \Comment{Compute join (locally on each node)}
  \ForAll{$node_j$} 
  \State $Result_j^V \leftarrow (\bigcup_{x=1}^{m}{d_{1xj}}) \Join \cdots \Join (\bigcup_{x=1}^{m}{d_{nxj}})$
\EndFor
\end{algorithmic}
\end{algorithm}

The  evaluation cost of the $Pjoin$ algorithm can be estimated as follows: 


\begin{eqnarray*}
Cost(Pjoin_V(q_1^{p_1}, \cdots , q_n^{p_n})) &=& \sum\limits_{1\leq i\leq n} Cost(q_i) \\
&& + \sum\limits_{1\leq i\leq n \land p_i \neq V}  Tr(q_i)
\end{eqnarray*}

where $Cost(q_i)$ is the cost of evaluating sub-query $q_i$, and
$Tr(q_i)$ is the cost for transferring (shuffling) the results $d_i$ that are not
yet partitioned on $V$.  The parameters for
estimating 
data transfer cost include the
result size 
\bernd{à vérifier: size vs. number of triple} 
 $\size(q)$ of each given sub-query $q$,
and the unit transfer cost $\ucom$.
$$Tr(q_i) = \ucom\times \size(q_i)$$ 
Notice that the cost of $Pjoin$ does not depend on the order of the $q_i$ sub-queries. 

For example, consider the sub-query $join_y(t_4, t_2, t_3)$ of $Q_8$
and the triple set $D$ is already partitioned by subject (for triple
patterns $t_2$ and $t_4$). An evaluation of this sub-query is
$Pjoin_y(t_4^y, t_2^y, t_3^x)$ and only needs to partition and shuffle the result of $t_3$ before computing the join. 
The corresponding total evaluation cost is $3 \cdot \uacc \cdot
\size(D)$ for data access and only $\ucom \cdot \size(t_3)$ for data
transfer.
\begin{figure}[htbp]
\centering
\includegraphics[width=\linewidth]{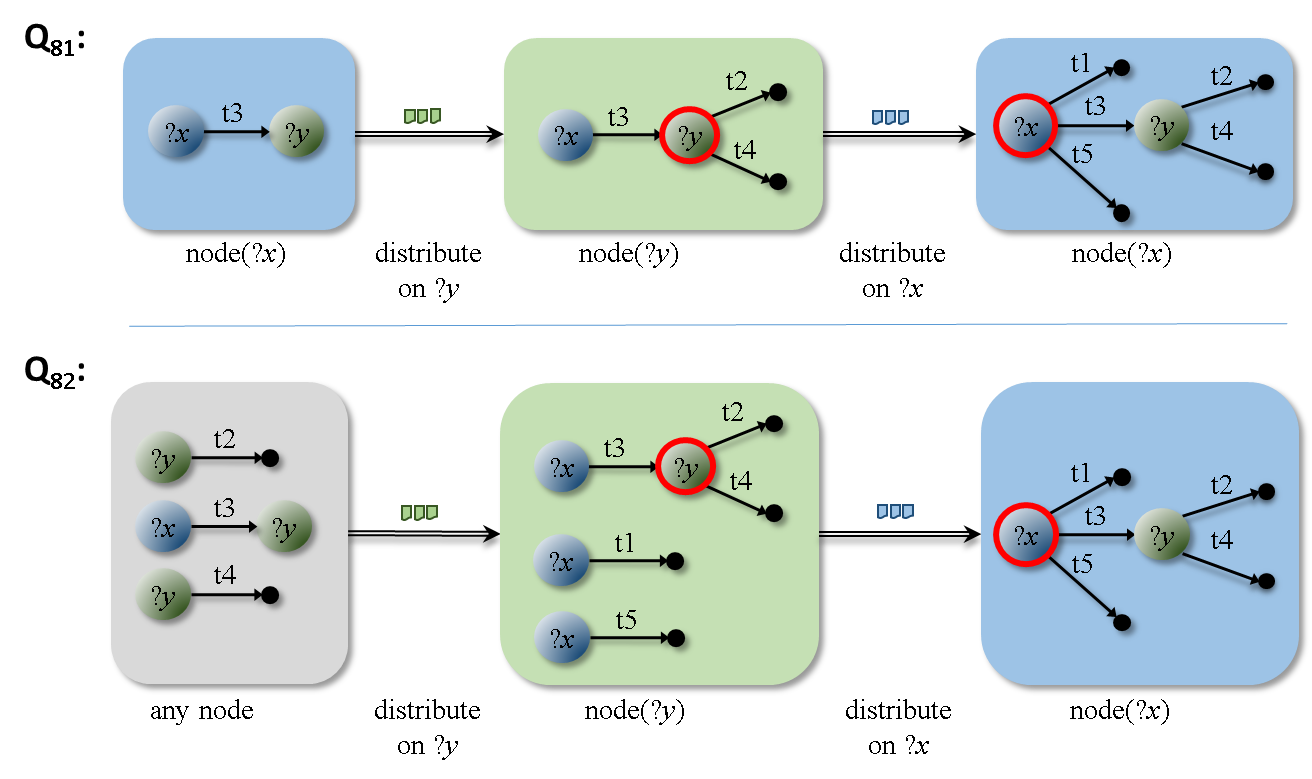}
\caption{Evaluation plans with partitioned join}
\label{fig:q81}
\end{figure}

%
%
%



\subsection{Broadcast Join: $Brjoin$} 
\label{sec:brjoin}
%

Broadcast join, denoted $q=Brjoin_V(q_1^{p_1}, \cdots, q_n^{p_n})$,
consists in sending the query results $d_i$ of \emph{all} sub-queries
except one, called the target query, to \emph{all} nodes
$node_j$. Without loss of generality, we assume that $q_n$ is the
target sub-query, excluded from the broadcast step, and has the
largest size among all sub-queries. The corresponding steps are
detailed in
Algorithm~\ref{alg:brj}. 






\begin{algorithm}
\caption{Broadcast Join} 
\label{alg:brj}
\begin{algorithmic}[1]
\State \textbf{Input}: $\{q_1^{p_1}, \cdots, q_n^{p_n}\}$, join variables $V$
\State \textbf{Output}: result fragment $Result_j$ on each node $node_j$
\Statex \Comment{Evaluate and broadcast sub-query results}
\ForAll{$q_i$ where $i<n$} 
   \ForAll{$node_j$} 
     \State { $d_{ij} \leftarrow$ evaluate $q_i$ on node $node_j$}
     \ForAll{$node_k\neq node_j$ }  
        \State{ transfer $d_{ij}$ from $node_j$ to $node_k$}
    \EndFor
  \EndFor
\EndFor
\Statex \Comment{Compute join (locally on each node)}
 \ForAll{$node_j$}   
   \State { $d_{nj} \leftarrow$ evaluate $q_n$ on node $node_j$}
   \State $Result_j \leftarrow \bigcup_{y=1}^{m}{d_{1y}} \Join \cdots \Join \bigcup_{y=1}^{m}{d_{(n-1)y}} \Join d_{n j}$
   \EndFor

\end{algorithmic}
\end{algorithm}




Broadcast join does not require any specific data partitioning and
preserves the partitioning of the target query, \ie the result of the
broadcast join has the same partitioning as the target query,
$q^{p_n}$.
The corresponding query cost is:

\begin{eqnarray*}
Cost(Brjoin_V(q_1, \cdots, q_n)) &=& \sum\limits_{i=1}^{n} Cost(q_i) \\
&&+ (m-1)\times \sum\limits_{i=1}^{n-1} Tr(q_i)
\end{eqnarray*}
where $Cost(q_i)$ and $Tr(q_i)$ are defined as before.
%
\begin{figure*}[htbp]
\centering
\includegraphics[width=0.8\linewidth]{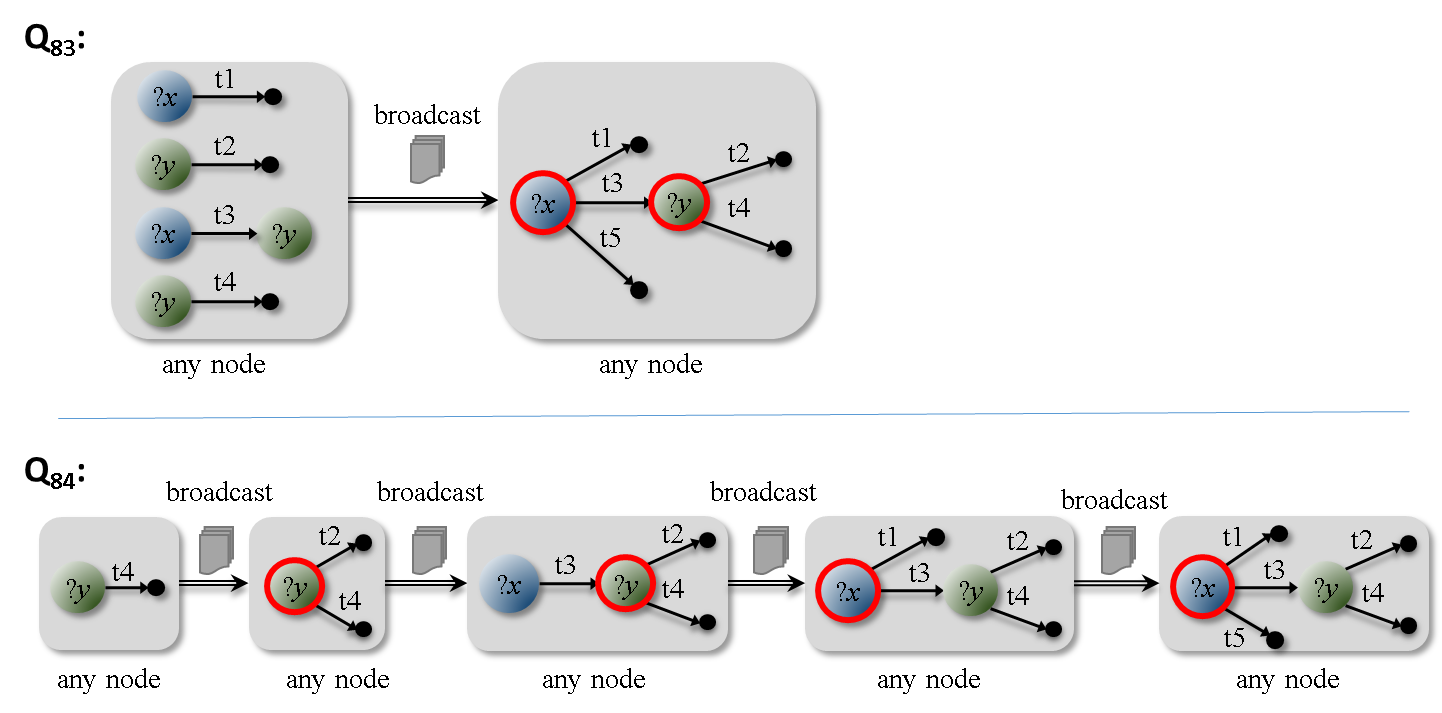}
\caption{Evaluation plans with broadcast join}
\label{fig:q83}
\end{figure*}
\sloppypar
For example, an evaluation of the query $join_y(t_4,t_2,t_3)$ is $Brjoin_y(t_4,t_2,t_3)$ which broadcasts the bindings of $t_4$ and $t_2$ to all nodes and then evaluates the join locally at each node.
Note that local join computation is possible whatever is the partitioning of $t_3$ since each node received a full replica of $t_2$ and $t_4$ during the broadcast step. 
The corresponding cost is $3 \cdot \uacc \cdot \size(D)$ for data access and  $(m-1) \cdot \ucom \cdot (\size(t_4) + \size(t_2))$ for data transfer.









\subsection{Building join plans} 
\label{sec:plan}

In this section we focus on the join optimization phase, and in
particular on the translation of a logical join tree into a physical plan.
We present four translation strategies which differ in their join
ordering capabilities and in the variety of physical join algorithms
they may use.  The first two translation
strategies are worth mentioning because they are  used in many
state-of-the art distributed SPARQL processing solutions. We consider
them as baselines. The last two methods are part of our contributions.
We illustrate each method on query $Q_8$  and its
logical expression: $join_x(join_x(join_y(join_y(t4,t2),t3),t1),t5)$.

\paragraph{PJoin plan}
The aim of this join strategy is to efficiently process star pattern
queries using partitioned join $Pjoin$ where the input data is already
partitioned by the join variable. This avoids data shuffling as shown
in the cost function of $Pjoin$.  For example, in our
experimentations, where all data is partitioned on the triple
subjects, this join is efficient on directed star queries joining
triples by their subject.  

This strategy translates each join into a $PJoin$ operator.  It also
merges two successive joins on the same variable into one n-ary
$Pjoin$. For instance $ Pjoin_{y}(Pjoin_{y}(t_4^y, t_2^y), t_3^x)$
becomes $Pjoin_{y}(t_4^y, t_2^y, t_3^x)$.  Figure~\ref{fig:q81} shows
the \textit{PJoin plan} of $Q_8$, over a triple set $D^s$ partitioned
by subject: $Q_{8_1} = Pjoin_{x}(Pjoin_{y}(t_4^y, t_2^y, t_3^x),
t_1^x, t_5^x)$. It distributes $t_3$ triples
(based on $y$) then joins them with $t_4$, $t_2$ on $y$. The result is
partitioned on $x$ and distributed to be joined with $t_1$ and $t_5$ on
$x$. The overall transfer cost of $Q_{8_1}$ is $\ucom \cdot
(\size(t_3) + \size( join_y(t_4,t_2,t_3)))$.

If $D$ is randomly partitioned, all sub-queries generate data
transfers between all nodes and we would get a much less efficient
plan $Q_{8_2}$: $Pjoin_{x}(Pjoin_{y}(t_3^\emptyset, t_2^\emptyset,
t_4^\emptyset), t_1^\emptyset, t_5^\emptyset)$ which, in comparison
with $Q_{8_1}$, adds the cost to transfer the results of $t_2, t_4,
t_1$, and $t_5$.

\paragraph{Mono Brjoin plan}
This strategy translates a join query into a single Brjoin
operator. It broadcasts all the operands of all the joins of the
query, except the last operand of the last join which is the target
query.  Figure~\ref{fig:q83} shows the \textit{Mono Brjoin plan} of
$Q_8$.  $Q_{8_3} = Brjoin_{yx}(t_4, t_2, t_3, t_1, t_5)$ where $t_5$
is the target sub-query.  This 5-way join plan implies to broadcast
the results of $t_4$, $t_2$, $t_3$, and $t_1$. The corresponding
transfer cost is $(m-1) \cdot \ucom \cdot [\size(t_4) + \size(t_2) +
\size(t_3)+ \size(t_1)]$.  Whereas this is less efficient than the
$Pjoin$ plan presented previously, it might become more efficient when the input
data is not partitioned. The broadcast join is also particularly
efficient when joining a small data set with a large one which is not
transferred.

%
%

\paragraph{Multi Brjoin plan}
This strategy considers only binary joins and aims to successively transfer
the smallest data set which may be the intermediary result of a join.
To this end, the \textit{Multi Brjoin} planning method expands the
query into a tree of binary joins, then translates each join into a
$Brjoin$ operator. Applied on  query $Q_8$, the resulting plan corresponds to $Q_{8_4}$ (see Figure~\ref{fig:q83}): $Q_{8_4} = Brjoin_x(Brjoin_x(tmp, t_1), t_5)$ with $tmp = Brjoin_y(Brjoin_y(t_4,
t_2), t_3)$. The
cost of $Q_{8_4}$ is $(m-1) \cdot \ucom \cdot [ \size(join(t_4) +
\size(join(t_4, t_2)) + \size(join(join(t_4, t_2),t_3)) +
\size(join(join(join(t_4, t_2),t_3),t_1)) ]$.  It is smaller than the
cost of $Q_{8_3}$ because the successive joins are rather selective
(\ie small result size).

\paragraph{Hybrid join plan}
We propose a cost-based join strategy which dynamically selects (i) the triple patterns that need to be joined and (ii) decides which join algorithm ($BrJoin$ and $PJoin$) is the most appropriate.
More precisely, given a logical query pattern $q=(t1, \cdots , t_n)$, the idea is to iteratively construct a linear binary join
tree where each step consists in choosing the next triple pattern and
join algorithm which generates the minimal cost. To estimate the size
of intermediate results, without any prior knowledge on the triples
pattern selectivity, we collect the size information at run-time as
suggested in~\cite{SIGMOD13:Shark}.
The building process starts by choosing two triple patterns $t_i$ and $t_j$ and a
join operator which generates the smallest cost. In general, when the data set is partitioned, this will
result in a partitioned join which is executed locally without any data transfer. The resulting plan $P_{ij}$
is extended as follows: for each $t_k$ that shares some join variables
with $P_{ij}$, estimate the cost values $Cost(PJoin(P_{ij}, t_k))$, $Cost(BrJoin(P_{ij}, t_k))$ and
$Cost(BrJoin(t_k, P_{ij}))$. Choose $t_k$ and its associated join
algorithm with minimum cost, and update $P_{ij}$ consequently. Repeat
this step with the remaining triple patterns until the entire plan
$P$ is built. The cost of this strategy is polynomial in the number of
triple patterns. There obviously exist join plan generation strategies
which might lead to more efficient) plans using more sophisticated
cost-estimation models and search algorithms. The goal of
this article is to show the benefit of generating hybrid join plans
using different physical join algorithms and exploring more
sophisticated join plan generation methods is part of our future work.
%

As emphasized in our evaluation (see Section \ref{eval}), this hybrid join planning strategy
allows for combining $PJoin$ and $BrJoin$ in a single query. Therefore, it supports snowflake queries efficiently because it generally ends up
with plans that locally evaluate  star sub-queries through n-ary $Pjoin$s
before joining these stars through a sequence of $BrJoin$ (assuming
selective stars, \ie the size of a star is always less than the total
size of the triple patterns that make the star).


Using our cost-model, we can easily show the advantage of our cost-based hybrid strategy.
Let $join_x(t_1, t_2)$, such that $\size(t_1) \leq \size(t_2)$ and $t_1$ and $t_2$ are not partitioned on $x$. Applying our cost model, we obtain the following three equivalent inequalities:
\begin{eqnarray*}
Cost(Pjoin_x(t_1^\emptyset, t_2^\emptyset)) &\leq& Cost(Brjoin_x(t_1, t_2))\\
\size(t_1) + \size(t_2) &\leq&  (m-1) . \size(t_1)\\  
 \frac{\size(t_2)}{\size(t_1)} + 2 &\leq& m
\end{eqnarray*}

The  choice of the join algorithm on two randomly distributed input data favors broadcast join  if the ratio of the input data size increases and partitioned join for an increasing number of machines $m$.

\bernd{j'ai avancé l'exemple ici:}

\begin{figure}[htbp]
\centering
\includegraphics[width=\linewidth]{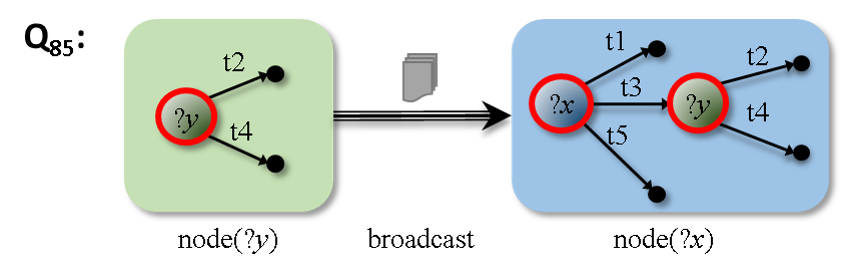}
\caption{Evaluation plan with Hybrid join plan}
\label{fig:q85}
\end{figure}

For example, consider $D^s$ partitioned on the triples subject. To evaluate $Q_8$ using a hybrid plan, first join $t_4$ with $t_2$ on $y$ without any transfer because $t_4$ and $t_2$ are adequately distributed on $y$. Then broadcast the result and join it on $x$ with the remaining patterns $t_3, t_1, t_5$ which are adequately distributed on $x$. We obtain the following expression (see Figure~\ref{fig:q85}):
$Q_{8_5} = Pjoin_x(Brjoin_y(Pjoin_y(t_4^y, t_2^y), t_3^x), t_1^x, t_5^x)$. The corresponding transfer cost is only $(m-1) \cdot \ucom \cdot \size(join(t_4, t_2))$ which is the smallest values compared to the plans generated by the other planning strategies.




Our hybrid planning strategy highlights that the total data transfer cost of a complex distributed join depends on the choice and the ordering of the joins. This query optimization problem has been studied in the context of massively distributed SPARQL query processing~\cite{DBLP:conf/icde/GoasdoueKMQZ15} and is complementary to our contribution.




\section{SPARQL Processing on Apache Spark}
\label{sec:query:spark}
We describe the main data abstraction models of the Apache Spark cluster computing platform. 
We propose five approaches to enable SPARQL query processing on top of Spark. 
We investigate how far each method supports the algorithms presented in Section~\ref{sec:sparql} to evaluate joins and generate join plans.





\subsection{Apache Spark}
\label{sec:spark}

Apache Spark \cite{DBLP:conf/hotcloud/ZahariaCFSS10} is a cluster computing engine and 
currently the big data-related Apache project with the most active contributors.  Spark can be apprehended as a main-memory extension of the MapReduce model enabling parallel computations on unreliable machines and automatic locality-aware scheduling, fault tolerance and load balancing. While both, Spark and Hadoop, are based on a data flow computation model, Spark is more efficient than Hadoop for applications requiring frequent reuse of working data sets across multiple parallel operations.  This efficiency is mainly due to two complementary distributed main-memory data abstractions: (i) Resilient Distributed Data sets (RDD) \cite{DBLP:conf/nsdi/ZahariaCDDMMFSS12}, a distributed, lineage supported fault tolerant memory data abstraction that enables one to perform in-memory computations (when Hadoop is mainly disk-based) and (ii)  Data Frames (DF), a compressed and schema-enabled data abstraction.
Both data abstractions ease the programming task by natively supporting a  subset of relational operators like $project$, $join$ and $filter$ which are not natively supported in Hadoop.  These operators enable the translation and processing of high-level query expressions (\eg SQL, SPARQL) using any programming language supported by Spark, \eg Java, Scala or Python as a host language.  
 
On top of RDD and DF, Spark proposes two higher-level data access models, GraphX and Spark SQL, for processing semi-structured data in general, \ie they can be adapted to handle RDF data and SPARQL queries.  Spark GraphX~\cite{DBLP:conf/osdi/GonzalezXDCFS14} is a library enabling the manipulation of graphs through an extension of Spark's RDD.  Being the most prone to represent graphical information, GraphX  seems at first sight to be a good candidate to implement a distributed SPARQL engine.  However, GraphX follows a vertex-centric computation model similar to Pregel~\cite{malewicz2010pregel} which is dedicated to perform highly-parallel graph algorithms such as  Pagerank and triangle counting. This processing model is not adapted to set-oriented graph pattern matching and is not considered in our proposed solutions.

Spark SQL~\cite{armbrust2015spark} allows for querying structured data stored in DFs. Its optimizer, Catalyst~\cite{armbrust2015spark}, is claimed to improve the execution of queries. This optimizer can also be used with the DF API which proposes a Domain Specific Language (DSL) to express queries. 

In the following Sections, we highlight how Spark SQL, RDD and DF can be used to develop different solutions for SPARQL query processing.

\subsection{SPARQL SQL}
\label{sec:query:sql}
The SPARQL SQL method consists in rewriting a given SPARQL query $Q$ into a SQL query $Q'$ which is evaluated by the Spark SQL engine~\cite{armbrust2015spark}. The execution plan of $Q'$ is determined by the embedded Catalyst optimizer using the Spark DF data abstraction which applies the broadcast join method. In our experiments with Spark SQL version 1.5.2, we observed that, when a query contains a chain of more than two triple patterns, a cartesian product is used rather than joins. Consider 3 triples patterns $t_1=(a,p1,x), t_2=(x,p2,y)$ and $t_3=(y,p3,b)$, and the query $join_y(join_x(t1,t2), t3)$. Then, for the corresponding SQL expression, Catalyst generates the physical plan $P = Brjoin_x(Brjoin_{-}(t_1, t_3), t_2)$ which computes a cross product between $t_1$ and $t_3$ before joining with $t_2$. This obviously is less efficient than, for example, plan $P'= Brjoin(Brjoin(t_1, t_2), t_3)$).
Plan $P$ in particular requires to broadcast the result of the Cartesian product $t1 \times t3$ whose size is very large compared to the size of $join(t1,t2)$. The cost of $P$ is:

$$Cost(P) = \sum\limits_{i=1}^{3} Cost(t_i) + (m-1) \cdot (Tr(t1) + Tr(t1 \times t3)))$$




This leads to very bad performance which is also confirmed by our experiments. We must underline that we did not explore this behavior in more detail and we are currently investigating the possibility of extending the Catalyst optimizer with appropriate declarative rewriting rules for achieving more efficient join plans~\cite{armbrust2015spark}.


\subsection{SPARQL RDD}
\label{sec:query:rdd}
The SPARQL RDD approach consists in using the Spark RDD data abstraction and specifically the $filter$ and $join$ methods of the RDD class for evaluating SPARQL queries over large triple sets.
Every logical join translates into a call to the $join$ method.
That $join$ method implements the $Pjoin$ algorithm which takes into account existing data partitioning to avoid data transfers whenever possible.
For example, the $Q_8$ query is evaluated according to the $Q_{8_1}$ plan.

SPARQL RDD evaluates the joins following the order specified by the input logical query, thus it conforms to the \textit{PJoin plan} strategy. 
SPARQL RDD does not perform any join optimization and ignores hybrid plans that may perform faster.
In particular, it misses an efficient solution when a broadcast join is cheaper (join a small with large data set).
Observe also that SPARQL RDD  always reads the entire data set for each evaluation of a triple pattern. We remedy this problem by merging multiple triple selections as described in~Section~\ref{sec:multisel}.


\subsection{SPARQL DF}
\label{sec:query:df}
Spark Data Frame (DF) provides an abstraction for manipulating tabular data through specific relational operators. Translating a SPARQL query using the DF DSL is straightforward:
triple selections translate into DF $where$ operators whereas SPARQL n-ary join expressions are transformed into trees of binary DF $join$ operators.
The main benefit of using this approach comes from the columnar, compressed in-memory representation of DF. The advantages are twofold. First it allows for managing larger data sets (\ie up to 10 times larger compared with RDD) for a given memory space, and second, DF compression saves data transfer cost.

DF uses a cost-based join optimization approach by preferring a single broadcast join (\ie mono $Brjoin$ plan defined in Section~\ref{sec:plan}) to a sequence of partitioned join (\ie $Pjoin$ plan) if the size of the data set is less than a given threshold. This achieves efficient query processing when joining several small data sets with a large one.

However we could observe two important drawbacks in applying the SPARQL DF approach.
The first drawback comes from the fact that DF only takes into account of the size of the input data set for choosing $Brjoin$. By this, DF does not efficiently handle very frequent join expressions $join(s, t)$ where $s$ is a highly selective filtering expression over a large data set. In that case, $Brjoin$ would be more efficient since it would avoid the data transfer for pattern $t$ (cost comparison for partitioned and broadcast joins in the section about \emph{Hybrid joins}).

The second drawback is that SPARQL DF (up to version 1.5) does not consider data partitioning and there is no way to declare that an attribute among ($s$, $p$ or $o$) is the partitioning key. Consequently, partitioned joins always distribute data and cause costly data transfers. This penalizes star pattern queries where the result of each triple pattern is already adequately distributed, since the query could have been answered without any transfer. 
Example $Q_{8_2}$ illustrates the processing of $Q_8$ through the DF layer. 



\subsection{SPARQL Hybrid execution}
\label{sec:query:hybrid}
The goal is to overcome the limitations found in the SPARQL RDD and SPARQL DF solutions in order to provide a more efficient SPARQL processing solution on top of Spark. 
%
In particular, we aim to: 
(i) take into account the current data partitioning to avoid useless data transfers, 
(ii) enable data compression provided by the DF layer to save data transfers and manage larger data sets, and 
(iii) reduce the data access cost of self-join operations.
To this end, the SPARQL Hybrid evaluation method implements two methods: the hybrid plan strategy and the merging of multiple triple selection proposed in Sections~\ref{sec:plan} and~\ref{sec:multisel} respectively.

Implementing the hybrid plan strategy requires a fine-grain control of the query evaluation plan at the operator level: each time the hybrid strategy chooses a join algorithm (say $J$), it runs a Spark job for $J$. Then, we can obtain the size of $J$ that is a required information for choosing the next join algorithm.

We propose a hybrid execution solution for both, RDD and DF data abstractions. 
\begin{itemize}
\item For the RDD abstraction (that does not support the $Brjoin$ natively) we decompose the $Brjoin$ operator into two jobs, one for broadcasting the data and the other for computing the join result based on the broadcast data using the mapPartitions method of the RDD API. 
\item For the DF abstraction, to ensure that the $BrJoin$ operator runs consistently according to the hybrid choice, we had to switch-off the less efficient threshold-based choice of the Catalyst optimizer.
\end{itemize}

Finally, to merge multiple triple selections, we add a preliminary step that persists the $S$ subset in main memory.



%

\subsection{Qualitative Comparison}
\label{sec:comp}
We have proposed an enhanced query execution method which allows for transferring only the necessary compressed data across the cluster' nodes, and permits to execute any query portion scanning the whole data set only once.
To better highlight the advantages of our method with respect to related ones, Table~\ref{tab:recap} presents a synthetic view of the query processing properties of all methods presented in this section. The different dimensions are:
\begin{itemize}
\item \textbf{Co-partitioning}. Triples that are partitioned on the join key can be joined locally without any transfer. DF does not support co-partitioning until Apache Spark version 1.5.

\item \textbf{Join algorithm}. Partitioned join $Pjoin$, $Brjoin1$ means that the support is restricted to a single $Brjoin$ per query, $Brjoin+$ means full support of several $Brjoin$ per query.

\item \textbf{Merged access}. The ability to evaluate several triple selections with a single scan of the data set.

\item \textbf{Optimization}. Ability to choose among several join algorithms. 
We qualify the choice done by Spark DF  as \enquote{poor} because it ignores broadcast join for large data set with highly selective triples patterns.

\item \textbf{Data compression}. DF abstraction uses data compression and allows for managing ten times larger data sets than RDD, at equal memory capacity.
\end{itemize}

The table shows that our \textit{SPARQL Hybrid} method offers equal or higher support for all the considered properties. Interestingly, \textit{SPARQL Hybrid} suits to both data abstractions, RDD and DF, because we strive to design our solution in a generic way decoupling the join optimization logic from the lower level Spark data representations.
We therefore are also confident that \textit{SPARQL Hybrid} could easily be extended to support forthcoming Spark data abstractions such as DataSet or GraphFrame.

\begin{table*}[htbp]
\begin{center}
\begin{tabular}{|l||c|c|c|c|c|}
\hline
                & \textbf{Co-}          &  \textbf{Join}    & \textbf{Merged} &  \textbf{Query }   & \textbf{Data} \\
\textbf{Method} & \textbf{partitioning} & \textbf{algorithm} & \textbf{access}     & \textbf{ optimization} & \textbf{compression}   \\
\hline
SPARQL RDD  & \cmark  & $Pjoin$                   & \xmark & \xmark        & \xmark  \\ 
\hline
SPARQL DF  & \xmark ($\le$ v1.5) & $Pjoin$, $Brjoin1$ & \xmark & poor          & \cmark \\
\hline
SPARQL SQL  & \xmark  & $Pjoin$, $Brjoin1$           & \xmark & cross-product & \cmark\\ 
\hline
SPARQL Hybrid RDD & \cmark  &  $Pjoin$, $Brjoin+$         & \cmark & cost-based    & \xmark \\
\hline
SPARQL Hybrid DF  & \cmark  & $Pjoin$, $Brjoin+$           & \cmark & cost-based    & \cmark \\
\hline
\end{tabular}
\caption{Qualitative analysis of 5 query processing methods}\label{tab:summary}
\label{tab:recap}
\end{center}
\end{table*}

\reb{On relying on an optimized SQL engine for processing SPARQL queries (Rev4): SparkSQL and DataFrames do provide optimization concerning join processing (e.g., choosing broadcast vs. partitioned join) however they do not address dedicated optimization for self joins.}




\section{Related Works}

In this section, we consider two kinds of work related to this paper's topic: distributed RDF database management systems (mainly concentrating on implementations based on the Hadoop ecosystem) and distributed join processing.

\subsection{Cluster-based RDF data stores}

\hub{voir S2RDF, Impala : les 2 sont "on top of an SQL engine"
Un moteur SQL ne peut traiter des workflows complexes avec des tâches d'analyse de graphe difficile/impossible à exprimer en SQL (modèle BSP de Pregel vs. SQL ) ?}

\reb{
Experimental comparison of our approach with other state-of-the art systems and infrastructures (Rev2,3,4) : our goal was to design an efficient distributed, main-memory triple store, without high indexing overhead, for improving scalability by reducing data preprocessing cost and global storage cost. Systems based on sophisticated data indexing approaches are not in the same category and we’ll try to clarify this in related work section.
}

\bernd{we should maybe add a short sentence telling that there are other distributed SPARQL query processing approaches (Ioana?, Mendelzon?) which are less related to our context.}
\bernd{we also should conclude this section by a general comparison with our approach.}

We are particularly interested in cluster-based solutions for the distributed evaluation of SPARQL expressions and in the following we will present the main representative approaches using the Apache Hadoop framework. 

SHARD \cite{DBLP:conf/oopsla/RohloffS10}  is the first effort to store RDF data on top of the Apache Hadoop distributed data storage (HDFS) and processing (MapReduce) software stack. Triple sets are stored in HDFS as flat files where each line represents all the triples associated with a given subject. The SHARD query processing engine iterates over the set of triples for each triple pattern in the SPARQL query, and incrementally attempts to bind query variables to literals in the triple data, while satisfying all of the query constraints.  The SPARQL query clauses are processed in several MapReduce steps, forcing high latency due to a large number of I/O operations over HDFS.

Huang et.al. \cite{DBLP:journals/pvldb/HuangAR11} describes an architecture which we denote by nHopDB and which is composed of a graph partitioner and a set of workers implemented by RDF-3X \cite{DBLP:journals/vldb/NeumannW10} database instances. RDF graphs (triple sets) are partitioned using a traditional graph partitioning algorithm (METIS) extended by a simple replication strategy for reducing costly data exchange between nodes. This allows certain queries to be executed locally on a single node and completely take benefit of the high-performance RDF-3X query engine.  Hadoop is used to supervise query processing of queries where the answer set spans multiple partitions. In this case, nHopDB suffers from Hadoop's start-up overhead and inherent I/O latencies, due to costly disk access operations.


The design of the SHAPE system~\cite{DBLP:journals/pvldb/LeeL13} is motivated by the limited scalability of graph partitioning-based approaches and applies simple hash partitioning for distributing RDF triples.  Like in nHopDB, SHAPE replicates data for achieving n-hop locality guarantees and takes the risk of costly inter-partition communication for query chains which are longer than their n-hop guarantee.

Sempala\cite{DBLP:conf/semweb/SchatzlePNL14} and S2RDF\cite{s2rdf} are recent systems that are also processing SPARQL queries on top a cluster computing engine. Sempala executes SPARQL queries over Impala's query processor. Impala\cite{DBLP:conf/cidr/KornackerBBBCCE15} is a Massively Parallel Processing (MPP) database on Hadoop that plans the parallelization and fragmentation of SQL queries using a dedicated query processor. In that context, Sempala is responsible for translating SPARQL queries into Impala's SQL dialect (mostly compliant with SQL92). In Sempala, the data is persisted in HDFS using Parquet's column store approach and Snappy's compression. The data layout of Sempala corresponds to a triples table (justified by triple patterns with unbound properties) along a so-called unified property table, \ie a unique relation composed of one column containing subjects and as many columns as there are properties. The overhead of this data layout is mitigated by the efficient representation of NULL values in Parquet. The unified property table layout is efficient for star-shaped queries but is not adapted to other query shapes. 

The on-disk storage approach of Sempala motivated its authors to  propose a new system called S2RDF\cite{s2rdf}. S2RDF is built on Spark and uses its SQL interface to execute SPARQL queries. Its main goal is to address efficiently all SPARQL query shapes. Its data layout corresponds to the vertical partitioning approach presented in \cite{DBLP:conf/vldb/AbadiMMH07}, \ie triples are distributed in relations of two columns (one for the the subject and one for the object) corresponding to RDF properties. So-called ExtVP relations are computed at data load-time using semi-joins, to limit the number of comparisons when joining triple patterns. Considering query processing, each triple pattern of a query is translated into a single SQL query and the query  performance is optimized using the set of statistics and additional data structures computed during this pre-processing step.  The data pre-processing step generates an important data loading overhead  which might be up to 2 orders of magnitude larger than the subject-based partitioning without replication as proposed in our solution.   
For lack of space, we refer to \cite{DBLP:journals/vldb/KaoudiM15} for a detailed presentation and complete comparison of other  RDF stores equipped with distributed SPARQL query processors. 

\subsection{Distributed multi-way join processing}
Distributed multiway join processing in general has been the topic of many research papers since decades~\cite{lu1991optimization} and we will cite only some more recent representative contributions parallel distributed multiway joins over partitioned data. In \cite{DBLP:conf/edbt/AfratiU10}, a solution is presented for the computation of multi-join queries in a single communication round. The algorithm was originally designed for the MapReduce approach, thus justifying the importance of limiting communication costs which are associated to a high IO costs. 
The authors of \cite{DBLP:conf/pods/BeameKS14} have generalized this single-communication n-ary join problem over a fixed number of servers and designed a new algorithm named HyperCube by providing lower and upper communication bounds. HyperCube is also at the origin of an implementation presented in~\cite{DBLP:conf/sigmod/ChuBS15}. This work is a very promising approach for evaluating SPARQL queries in a MapReduce setting where the number of rounds has to be restricted. We have chosen a different setting, where data is in main memory and already partitioned with the goal to reduce the whole data transfer cost independently of the number of rounds (join tree depth). Nevertheless, we believe that it might be interesting to study the benefit of using Hypercube joins in our setting.


\hub{Qu'est-ce qu'on fait avec cette ref ? Join optimization\cite{DBLP:conf/edbt/TsialiamanisSFCB12})}


\section{Experimental Evaluation}
\label{eval}


\subsection{Experimental Setting}
The evaluation was conducted on a cluster of 18 DELL PowerEdge R410 running a Debian GNU/Linux distribution with a 3.16.0-4-amd64 kernel version. Each machine has 64GB of DDR3 RAM, a 900GB 
disk and two Intel Xeon E5645 processors. Each processor is constituted of 6 cores running at 2.40GHz and allowing to run two threads in parallel (hyper threading). 
The machines are connected via a 1GB/s Ethernet network adapter.
We used Spark version 1.6.2 and implemented all experiments in Scala. 
The Spark configuration of our evaluation runs our prototype on a subset of the cluster corresponding to 300 cores and 50GB of RAM per machine.

\reb{SparkSQL: We based our SPARQL to SQL translation (Rev3,4) using a standard approach (Sparqlify). This translation is quite straightforward, since we consider only graph patterns without filtering, union and optional. The translation into RDD and DF uses a Sesame-based parsing solution for the creation of logical/physical plans and code generation. All these steps are implemented in Scala. Experimentation reproducibility: by conference start, the code will be pushed on github.}

\paragraph{Data and Query Workload}
\label{sec:workload}
We have selected two synthetic and three real world knowledge bases. 
The synthetic data sets come from well-established Semantic Web benchmarks: the LeHigh University Benchmark (LUBM) and the Waterloo SPARQL Diversity Test Suite (WatDiv)~\cite{iswc:watdiv:2014}. The two benchmarks provide an ontology, a data generator, and a set of queries.
The LUBM data sets \cite{DBLP:journals/ws/GuoPH05} respectively store over 100 millions and one billion triples, resp. denoted LU100M and LU1B. We also used the same one billion triples WatDiv data set as in~\cite{s2rdf} for comparison purposes.
The real world data sets correspond to open source DBPedia, Wikidata and DrugBank RDF dumps.
The main characteristics of the data sets are reported in the first column of  Table~\ref{compression}.
We will validate the query processing methods presented in Section~\ref{sec:query:spark} over three common SPARQL query shapes, \ie star, chain, and snowflake. We consider that these three forms can be easily combined to describe other query shapes, \eg triangles, trees. 
More details on the evaluation can be obtained on this paper's companion web site\footnote{https://sites.google.com/site/sparqlspark/home}.

\subsection{Impact of Compression on Storage and Access Time}
Data compression is appealing to manage larger data sets for a given amount of memory.
We quantify the space benefit of compression according to the penalty for storing and accessing compressed data.
Table~\ref{compression} emphasizes the size of the uncompressed RDD and the compressed DF representations as well as the overheads to store and access the triples.
We report the absolute increase in time due to data compression (DF creation time - RDD creation time) and access/decompression (DF scan time - RDD scan time).
The table highlights that DFs occupies a much smaller space than RDD with a compression rate  between 8 and 17\%. 
The high compression rate comes at a low pre-processing cost. In fact, for the largest data set, creating a compressed DF from an existing uncompressed RDD only adds an overhead of less than 4 seconds (\ie 27ms/GB). 
%
The absolute access time overhead remains below 1 sec (\ie 804ms max) and has a minor impact when querying large data sets compared to the data transfer costs. 
For our largest data set containing 1.3 billion triples, it took on average 950ms to read the entire RDD data set vs. 1700 ms for the corresponding DF data set.

\begin{table*}[htbp]
\centering
\begin{tabular}{|c||r|r|r|r|r|r|}
\hline
                 & \# triples     & RDD size & DF size & Size ratio & Compress. time  & Access time \\
\textbf{Data set} &  $\times 10^6$ & (GB)    & (GB)  &   (in \%)  &  overhead (ms) & overhead (ms)\\
\hline
Drugbank & 0.505 & 0.275 & 0.022 & 8.0\%    & 500    & 202\\
\hline
DBpedia & 77.5 & 8.1 & 1.4 & 17.2\% & 290   & 213 \\
\hline
LU100M  & 133.5 & 13.9  & 1.58 & 11.4\%       &   250   & 320\\
\hline
Wikidata & 233.1&24.3 & 2.6 &   10.7\%   &  1550    & 293\\
\hline
LU1B & 1334.7 & 139.2 & 18.7 & 13.4\%    & 3850  & 804\\
\hline
\end{tabular}
\caption{Data sets and compression rates}
\label{compression}
\end{table*}
\subsection{Query Processing Performance}
\label{qpp}
We compare the performance of five query processing methods over oriented star, chain and snowflake queries.
We adopt a simple partitioning strategy (\cf Section~\ref{sec:datapart}) where all data sets are partitioned on their triple subject nodes.
Moreover, we compare our solution with the S2RDF~\cite{s2rdf} related one.

\reb{we will also will add additional figures about data processing and data exchange costs}


\subsubsection{Star Queries} 

\begin{figure*}[htbp]
\centering
\subfigure[Star queries]{\includegraphics[width=0.38\linewidth]{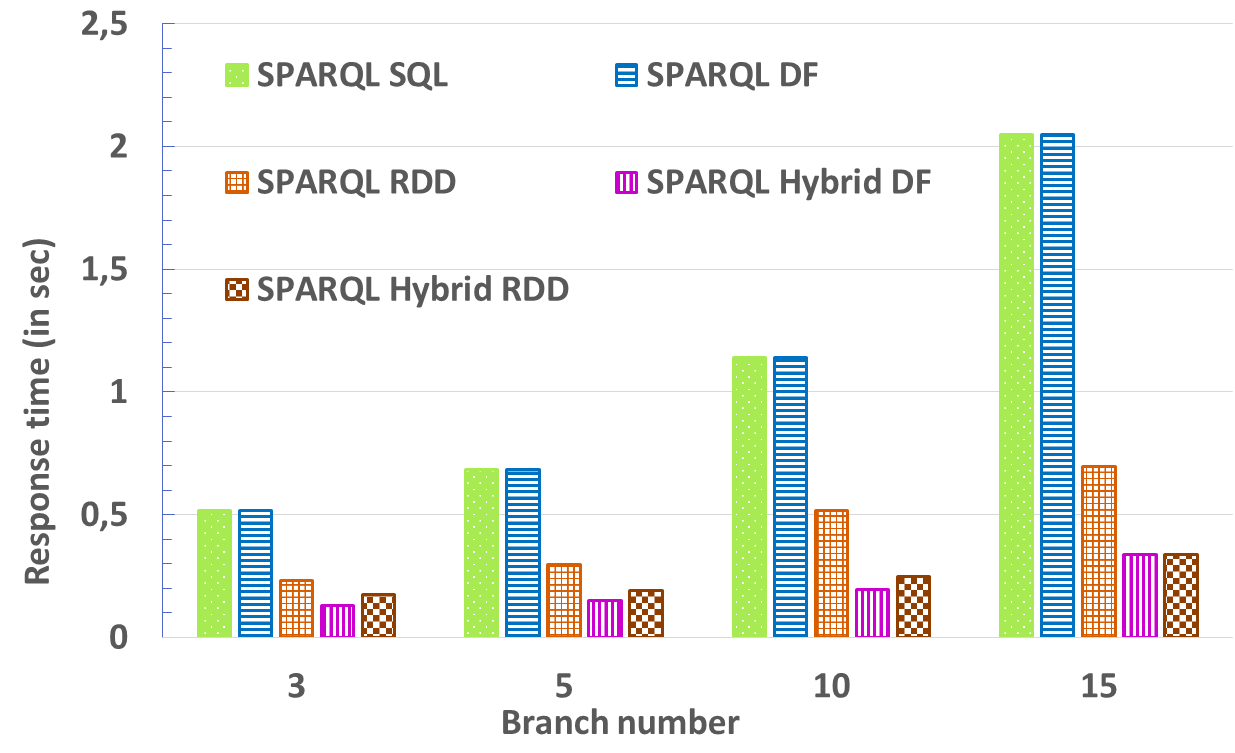}}\hspace{2cm}
\subfigure[Chain queries]{\includegraphics[width=0.38\linewidth]{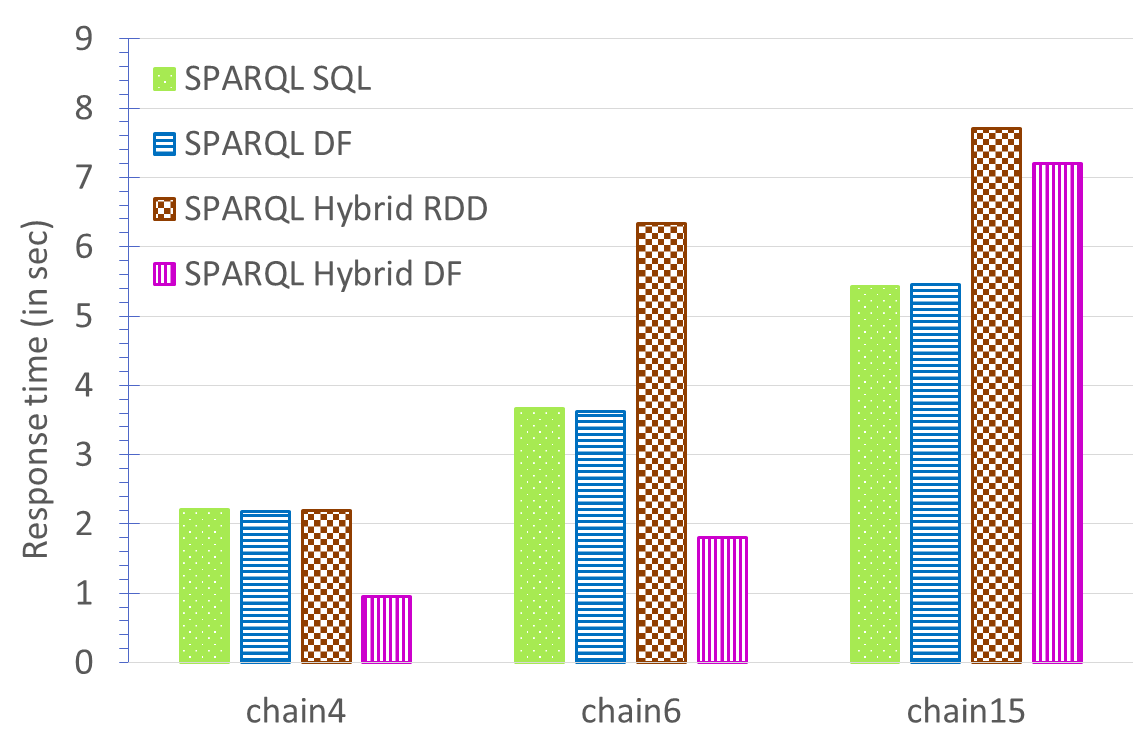}}
\caption{Query response time wrt. evaluation strategies on real world data sets}
\label{fig:starchain}
\end{figure*}

This experiment is conducted over the DrugBank knowledge base which contains high out-degree nodes describing drugs.
A first practical use case is to search for a drug satisfying multi-dimensional criteria and we defined four star queries with a number of branches ranging from 3 to 15.
We process each query using our five SPARQL query processing approaches and report query response times in Figure~\ref{fig:starchain}(a).
SPARQL SQL decides to reorder the joins only if it reduces the number of join operations which is obviously not possible for a star query containing only one join variable. Thus, SPARQL SQL generates the same evaluation plan (and cost) as SPARQL DF. Both methods ignore the actual data partitioning and broadcast the result of every triple pattern across the machines. On the opposite, SPARQL RDD, SPARQL Hybrid RDD, and SPARQL Hybrid DF are aware that the data are partitioned on the subject (\ie the join variable) and thus decide to process the query without any data transfer. Observe that the total costs is dominated by the transfer cost which explains why SPARQL DF is at least 2.2 times slower than the transfer-free methods.

\begin{figure*}[htbp]
\centering
\subfigure[LUBM query Q8]{\includegraphics[width=0.38\linewidth]{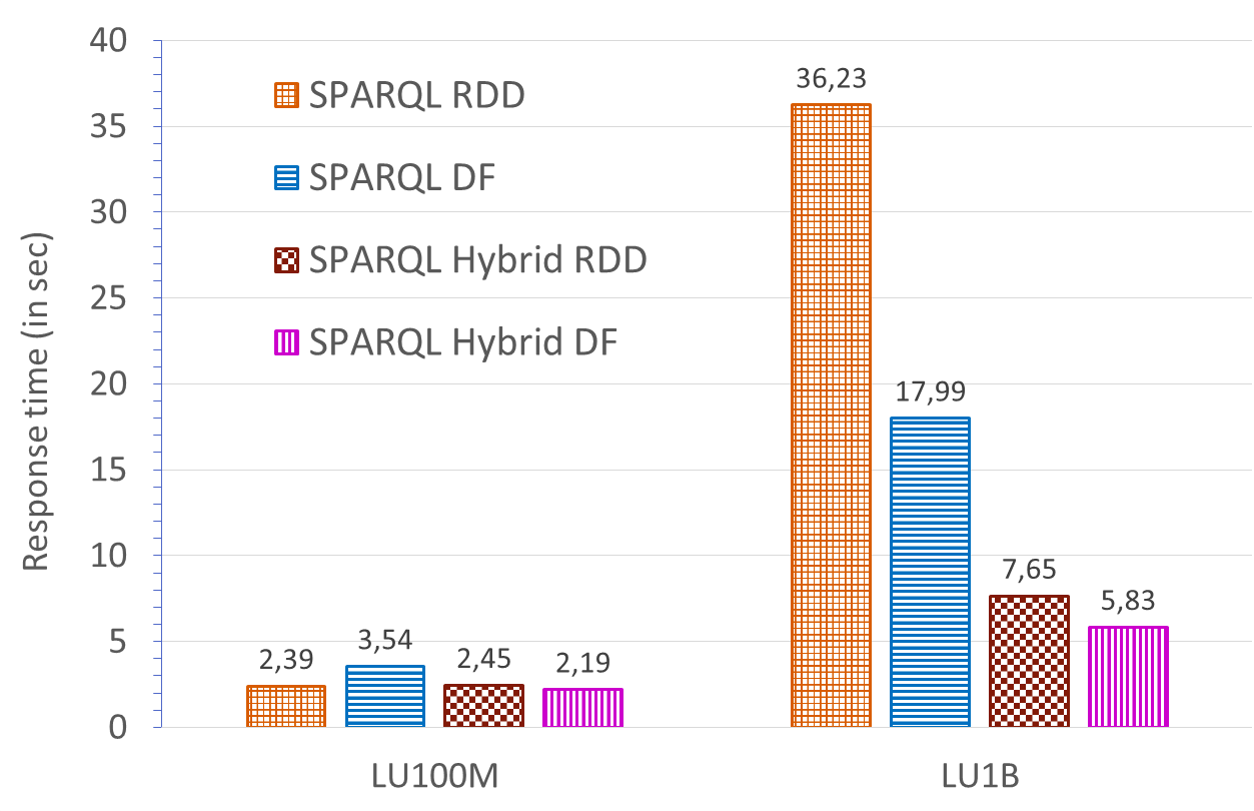}}\hspace{2cm}
\subfigure[WatDiv queries on 1B triples]{\includegraphics[width=0.38\linewidth]{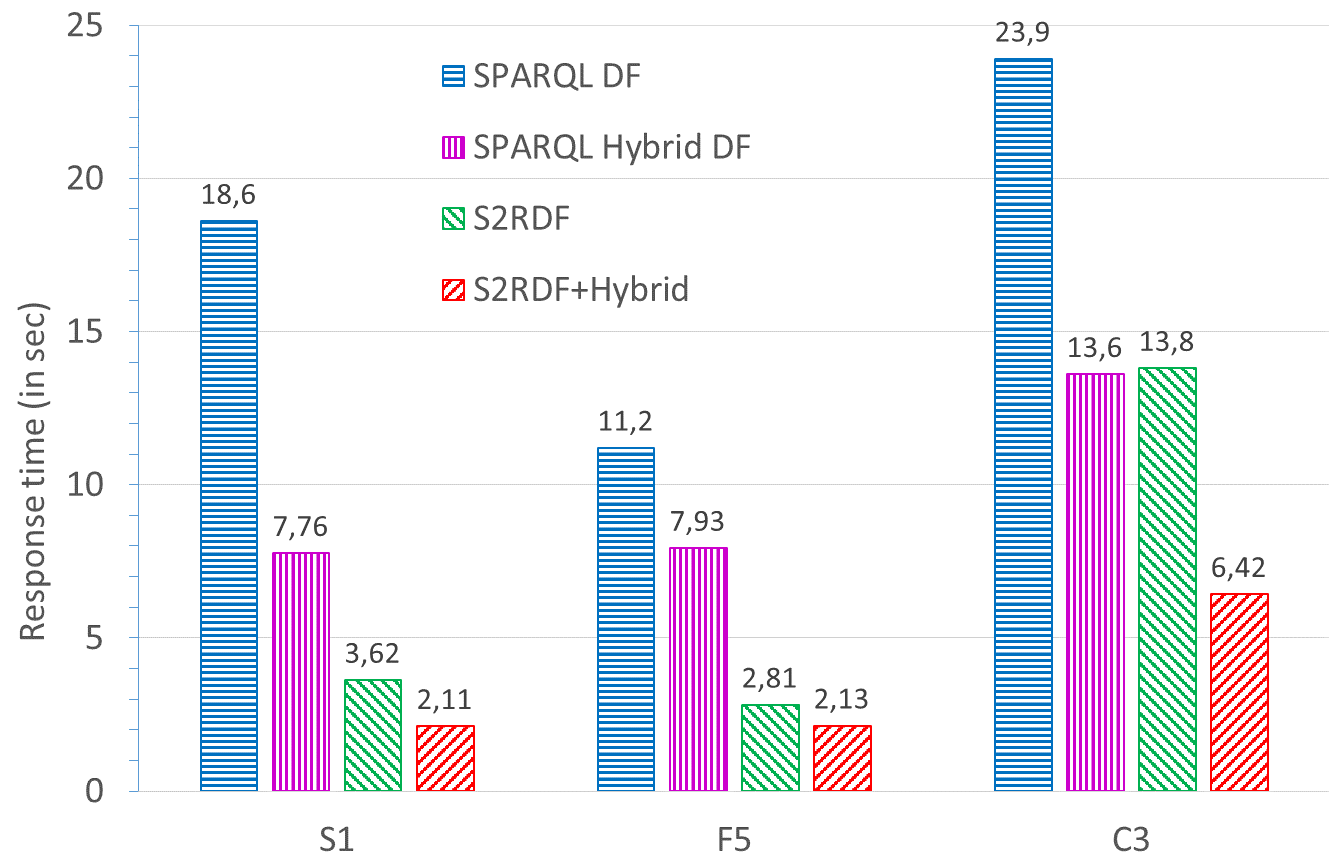}}
\caption{Performance of benchmark queries and comparison with S2RDF}
\label{fig:snow}
\end{figure*}

When comparing the transfer-free methods, we observe that both SPARQL Hybrid methods are 1.4 to 2 times faster than SPARQL RDD. In fact, SPARQL Hybrid reads the data set only once, whereas the data access cost of SPARQL RDD is proportional to the number of branches (triple selection patterns). 
Finally, SPARQL Hybrid DF slightly outperforms SPARQL Hybrid RDD for star queries with up to 10 branches. This is due to the way SPARQL Hybrid  implements partitioning of intermediate results: SPARQL Hybrid RDD relies on a built-in \texttt{groupBy} operator whose implementation is slightly more efficient than the user defined \texttt{groupBy} operator of SPARQL Hybrid DF.

\subsubsection{Property Chain Queries}
This experiment is done over the DBPedia knowledge base and a set of queries with a path length ranging from 4 to 15.  We report the query response times in Figure~\ref{fig:starchain}(b).
%
We first consider property patterns alternating frequent and rare properties, \ie composed of a triple pattern with a frequent property followed by one with an infrequent property, and so on. Queries $chain4$ and $chain6$ belong to this category.
We observe that for these queries, SPARQL Hybrid DF outperforms the other approaches. 
The relatively poor performance of SPARQL DF is due to its inability to estimate the intermediary result set's size. In fact, SPARQL DF is not informed that some triple patterns are very selective (the result size of triple patterns ranges from one hundred to one million triples),
and that it would be worth using a broadcast join rather than a partitioned join.

Our SPARQL Hybrid methods estimate the size of intermediary results at run time. 
Based on that, SPARQL Hybrid chooses to broadcast the results of all the triples patterns except the largest one (target pattern), then evaluates the entire chain query without additional data transfers. 
%
Furthermore, as explained in Section~\ref{sec:query:hybrid}, SPARQL Hybrid is able to get all intermediary results by accessing the data set only once (\cf the merged access property). The experiments show that the overall performance gain (of the informed choice combined with the merged access) is about 2 times faster compared with the other methods. 

Our experiment confirms the benefit of choosing a broadcast join 
for a chain made of a large size triple pattern followed by a smaller one.
Indeed the benefit is increasing when size ratio gets larger.

%
Finally, we show the case of query $chain15$ for which SPARQL DF outperforms SPARQL Hybrid DF. In this specific query, the first triple patterns (say $t$) and the following one (say $t'$) are large compared to the other ones and joining $t$ with $t'$ yields a very small intermediate result. Thus, it is efficient to run a partitioned join which distributes the first triple and each consecutive intermediate result. That is SPARQL DF plan. In contrast, SPARQL Hybrid DF starts with the lowest cost broadcast join of two triple patterns different from $t$ and $t'$.

\subsubsection{Snowflake Queries}
First, we focus on the most complex snowflake query of the LUBM benchmark ($Q8$). The evaluation plans for $Q8$ have been introduced in Section~\ref{sec:sparql} and we report the response times in Figure~\ref{fig:snow}(a).

%
%
%
%

Q8 does not run to completion with SPARQL SQL. The evaluation plan contains a cartesian product that is prohibitively expensive. This emphasizes that the Spark's Catalyst optimizer strategy to replace two joins by one cartesian product should be applied more adequately by taking into account the actual transfer cost.
%
SPARQL DF and SPARQL RDD confirm that working with compressed data is beneficial as soon as the data set is large enough. Although SPARQL DF ignores data partitioning, thus distributing more triples (320M instead of 104M triples for the partitioning-aware approach), its transfer time is lower than SPARQL RDD, thanks to compression. 

The major experimental result is that SPARQL Hybrid  outperforms existing methods by a factor of 2.3 for compressed (DF) data and 6.2 for uncompressed (RDD) data.  This is mostly due to reduced transfers (only few hundred triples instead of over one hundred million triples for the best existing approach). SPARQL Hybrid  also saves on the number of data accesses: two against three and five for respectively SPARQL RDD and SPARQL DF.



\subsubsection{Comparison with S2RDF}
Finally, we compare our Hybrid approach with the state of the art S2RDF~\cite{s2rdf} solution which outperforms many existing distributed SPARQL processing solutions.
We conducted the S2RDF comparative experiments over the same WatDiv 1 billion triple data set on a cluster with approximately similar computing power than the one used in the S2RDF evaluation (we used 48 cores in our experiment against 50 cores used in the S2RDF experiments). 
Our main goal was to show that our solution is complementary and can be combined with the S2RDF approach. For this, as a baseline we first selected three representative queries from the WatDiv query set, one for each category: S1 is a star query, F5 a snowflake one, and C3 a complex one (\cf Appendix). We executed S1, F5 and C3 over one large data set containing all the triples (\ie without S2RDF VP fragmentation), using SPARQL SQL and SPARQL Hybrid strategies. Then, we split the data set according to the S2RDF VP approach
(\ie one data set per property) and ran the queries using SPARQL SQL along with the S2RDF ordering method, and SPARQL Hybrid strategies. Response times are reported in Figure~\ref{fig:snow}(b).

Our SPARQL Hybrid solution outperforms the baseline SPARQL SQL by a factor ranging in $[1.76, 2.4]$ and the S2RDF solution by a factor ranging in $[1.72, 2.16]$ which is encouraging. 
The benefit mainly comes from reduced data transfers: our approach saved 483MB for S1, 284MB for F5, and 1.7GB for C3.
Note that while reproducing the S2RDF experiments, we get response times more than twice faster than those reported in ~\cite{s2rdf} (\eg 3.6 sec instead of 8.8 sec for query S1) and our 1.72 minimal improvement ratio is a fair comparison.
%
This highlights that our approach easily combines with S2RDF to provide additional benefit. We did not compare our approach with the concept of ExtVP relations of S2RDF's solution since it comes at high pre-processing overhead (17 hours for pre-processing the 1 billion triple data set) which does not comply with our  objective of reducing data pre-processing and loading cost.

\hub{Comparison with S2RDF: we ran the F5 WatDiv query. We measured 2.1sec with SPARQL Hybrid }

\section{Conclusion}
In this paper, we present the first exhaustive study comparing SPARQL query processing strategies over an in-memory based cluster computing engine (Apache Spark).
Five SPARQL execution strategies have been implemented and evaluated over different query shapes and data sets. The results emphasize that hybrid query plans combining partitioned join and broadcast joins improve query performance in almost all cases. Although SPARQL Hybrid RDD is slightly more efficient than the hybrid DF solution due to the absence of a data compression/decompression overload, it becomes interesting to switch to the DF representation when the size of RDDs almost saturates the main-memory of the cluster. In this, one can store almost 10 times more data on the same cluster size with only a small loss in performance. As future work, we are planning to extend our prototype with a full-fledged query optimizer and to study the  benefit and eventual implementation of other more recent distributed join algorithms such as HyperCube.

\bibliographystyle{abbrv}
\bibliography{amw} 

\appendix
\label{app:watdiv}
\paragraph{WatDiv Queries}
\begin{verbatim}
S1: SELECT ?v0 ?v1 ?v3 ?v4 ?v5 ?v6 ?v7 ?v8 ?v9 WHERE {
	?v0	gr:includes	?v1 .    %v2%	gr:offers	?v0 .
	?v0	gr:price	?v3 .       ?v0	gr:serialNumber	?v4 .
	?v0	gr:validFrom	?v5 .   ?v0	gr:validThrough	?v6 .
	?v0	sorg:eligibleQuantity	?v7 .
	?v0	sorg:eligibleRegion	?v8 .
	?v0	sorg:priceValidUntil	?v9 .  }

F5: SELECT ?v0 ?v1 ?v3 ?v4 ?v5 ?v6 WHERE {
	?v0	gr:includes	?v1 .   %v2%	gr:offers	?v0 .
	?v0	gr:price	?v3 .      ?v0	gr:validThrough	?v4 .
	?v1	og:title	?v5 .      ?v1	rdf:type	?v6 .  }

C3: SELECT ?v0 WHERE {
	?v0	wsdbm:likes	?v1 .   ?v0	wsdbm:friendOf	?v2 .
?v0	dc:Location	?v3 .   ?v0	foaf:age	?v4 .
?v0	wsdbm:gender	?v5 .  ?v0	foaf:givenName	?v6 . }
\end{verbatim}

\end{document}